\documentclass[twocolumn]{aastex63}
\usepackage{newtxtext,newtxmath}

\newcommand{\MJ}{M_{\rm J}}
\newcommand{\RJ}{R_{\rm J}}

\newcommand{\obl}{\theta_{\rm SL}}

\newcommand{\dego}{^\circ}
\newcommand{\nhat}{\hat{\bf n}}
\newcommand{\vhat}{\hat{\bf v}_{\rm proj}}
\newcommand{\rhat}{\hat{\boldsymbol \rho}}
\newcommand{\rphat}{\hat{\bf r}_{\rm rel}}

\newcommand{\Sm}{{\bf S}}
\newcommand{\rrel}{{\bf r}_{\rm rel}}
\newcommand{\vrel}{{\bf v}_{\rm rel}}

\newcommand{\phir}{\phi_{\bf r}}
\newcommand{\thetar}{\theta_{\bf r}}

\received{XXXX}
\revised{XXXX}
\accepted{XXXX}
\submitjournal{ApJL}

\shorttitle{Planetary Spin and Obliquity from Mergers}
\shortauthors{Jiaru Li \& Dong Lai}
\graphicspath{{./}{figures/}}

\begin{document}

\title{Planetary Spin and Obliquity from Mergers}

\correspondingauthor{Jiaru Li}
\email{jiaru\textunderscore li@astro.cornell.edu}

\author[0000-0001-5550-7421]{Jiaru Li}
\affiliation{Center for Astrophysics and Planetary Science,
Department of Astronomy, Cornell University, Ithaca, NY 14853, USA}
\affiliation{Theoretical Division, Los Alamos National Laboratory, Los Alamos, NM 87545, USA}

\author[0000-0002-1934-6250]{Dong Lai}
\affiliation{Center for Astrophysics and Planetary Science,
Department of Astronomy, Cornell University, Ithaca, NY 14853, USA}



\begin{abstract}

In planetary systems with sufficiently small inter-planet spacing, close encounters can lead to planetary collisions/mergers or ejections.  We study the spin property of the merger products of two giant planets in a statistical manner using numerical simulations and analytical modeling. Planetary collisions lead to rapidly rotating objects and a broad range of obliquities. We find that, under typical conditions for two-planet scatterings, the distributions of spin magnitude $S$ and obliquity $\obl$ of the merger products have simple analytical forms: $f_{S} \propto S$ and $f_{\cos\obl} \propto (1-\cos^2\obl)^{-1/2}$. Though parameter studies, we determine the regime of validity for the analytical distributions of spin and obliquity. Since planetary mergers is a major outcome of planet-planet scatterings, observational search for the spin/obliquity signatures of exoplanets would provide important constraints on the dynamical history of planetary systems.

\end{abstract}



\section{Introduction}  \label{sec:intro}

It has long been recognized that planet-planet interactions play an important role in shaping the architecture of planetary systems. Close planet encounters can lead to violent outcomes such as planetary mergers and ejections of one of the encountering planets.

There exists a large literature on giant planet scatterings \citep[e.g.,][]{Chambers1996,Rasio1996,Lin1997,Ford2001,Adams2003,Chatterjee2008,Ford2008,Juric2008,Nagasawa2011,Petrovich2014,Anderson2020}. Most of these focus on ejections and using the remnants of scatterings to explain the eccentricity distribution of extrasolar giant planets.

In contrast, there has not been much discussion on the merger products. Numerical simulations indicate that the ratio of ejections to planet-planet collisions depends on the ``Safronov number'', the squared ratio of the escape velocity from the planetary surface to the planet's orbital velocity; when the Safronov number is less than unity, a significant fraction of planetary collisions are expected \citep[e.g.,][]{Ford2001,Petrovich2014}. A comprehensive study of scatterings in systems with three giant planets shows that the collision fraction increases from $50\%$ at 1 AU to more than $80\%$ at 0.1 AU \citep{Anderson2020}. Our recent study of two-planet scatterings, including hydrodynamical effects, shows that even at 10 AU, the collision fraction can reach $40\%$ \citep{Li2020}. 

Previous studies on the collisions between protoplanetary objects have aimed mainly at understanding the process of late bombardment, during which collisions could be highly hyperbolic and the reaccretion efficiency is uncertain \citep{Agnor2004,Asphaug2006,Leinhardt2011,Stewart2012}. However, for giant planet collisions resulting from orbital instabilities, the relative motions are close to parabolic \citep{Anderson2020}, and the planets merge without significant mass loss (\citealt{Li2020}; see also \citealt{Leinhardt2011}). This implies angular momentum conservation in the colliding ``binary'' planets for a wide range of impact parameters. 

In this paper, we study planetary spin and obliquity generated by giant planet collisions. It is well recognized that the spin of a planet (both magnitude and direction) may provide important clue to its dynamical history. Various mechanisms have been suggested to produce non-zero planetary obliquities \citep[e.g.,][]{Safronov1969,Benz1989,Korycansky1990,Tremaine1991,Dones1993,Lissauer1997,Ward2004,Hamilton2004,Morbidelli2012,Vokrouhlicky2015,Millholland2019,Rogoszinski2020,Su2020}. Despite the lack of direct measurement of extrasolar planetary spins and obliquities, constraints can be obtained using high-resolution spectroscopic observations \citep{Snellen2014,Bryan2017,Bryan2020}. High-precision photometry of transiting planets can also help constrain planetary rotations in the future \citep[e.g.,][]{Seager2002,Barnes2003,Schwartz2016}.

We carry out a suite of numerical experiments of two giant planet scatterings to determine the distributions of spin and obliquity of the merger products. Based on our recent work on the hydrodynamics of giant planet collisions \citep{Li2020}, we assume that two colliding planets always merge into a bigger one with no mass loss. The rest of this paper is organized as follows. In Section~\ref{sec:simulation}, we present our fiducial numerical simulations and results. We then provide a simple analytical model in Section~\ref{sec:analytical} to explain the numerical distributions of spin and obliquity. We examine the limitation of our analytical model using parameter studies in Section~\ref{sec:parameter} and conclude in Section~\ref{sec:summary}.

\section{Fiducial Numerical Experiments}
\label{sec:simulation}

\subsection{Set-up of the simulations and assumptions}
\label{sec:set-up}

We consider a systems of two planets with masses $m_1 = 2\MJ$, $m_2 =1\MJ$ and radii $R_1 = R_2 = \RJ$, orbiting a host star with mass $M_*= M_{\sun}$ and radius $R_{\sun}$. The initial spacing (in semi-major axis) of the planets is given by
\begin{equation}
    a_2 - a_1 = 2.5 R_{\rm H,mut},
\label{eq:a1a2}
\end{equation}
where $R_{\rm H,mut}$ is the mutual Hill radius
\begin{equation}
    R_{\rm H,mut} = \frac{a_2+a_1}{2}\left(\frac{m_1+m_2}{3M_*}\right)^{1/3}.
    \label{eq:RH}
\end{equation}
This spacing is smaller than the critial value ($2\sqrt{3}R_{\rm H,mut}$) for the Hill instability \citep{Gladman1993}. In our fiducial runs, we use $a_1=1$ AU. For each planet, we sample the initial eccentricity in the range $[0.01,0.05]$, the initial inclination in $[0^{\circ},2^{\circ}]$, and the argument of pericenter, longitude of ascending node, and mean anomaly in the range $[0,2\pi]$, assuming they all have uniform distributions.

The simulations are performed using the open-source N-body software package \texttt{REBOUND} \footnote{\texttt{REBOUND} is available at \url{http://github.com/hannorein/rebound}.}\citep{Rein2012}. We choose the \texttt{IAS15} integrator \citep{Rein2014} for high accuracy because the planets can have small separations. We run each simulation up to $10^5$ initial orbital periods of the inner planet, and stop the simulation whenever one of the following conditions is reached:
\begin{itemize}
\item Collision: The relative separation $|\rrel|$ of the planets is equal to the sum of their physical radii (i.e. $|\rrel|=R_1+R_2$).
\item Ejection: One of the planets reaches a distance of $1000$~AU from the system's center of mass.
\item Star-Grazing: The distance between the star and one of the planets is less than the solar radius.
\end{itemize}
We focus on collisions in this paper. We assume the two planets have a perfect merger with no mass and angular momentum loss -- This is justified by our
hydrodynamical simulations \citep{Li2020}. 

The initial (pre-merger) spin of each planet is unknown. The current 10-hr spin period of Jupiter and Saturn corresponds to $30\%$ of the break-up rotation rate $\Omega_{\rm break}=(GM_{\rm p}/R_{\rm p}^3)^{1/2}$. Recent constraints on the spin of young planetary-mass companions also suggest that similar sub-break-up rotations are common for extrasolar giant planets \citep{Bryan2017}. Such slow rotation rate may result from the magnetic disk braking during or immediately after the formation the planet \citep{Takata1996,Batygin2018,Ginzburg2019}. Adopting the moment of inertia $I\simeq 0.26 M_{\rm p}R_{\rm p}^2$ and initial spin $\Omega_i\sim 0.3\Omega_{\rm break}$, the initial spin angular momentum of each planet is $S_{\rm init}\sim 0.078 (GM_{\rm p}^3R_{\rm p})^{1/2}$. On the other hand, the relative orbital angular momentum $L_{\rm orb}$ of the two planets just before merger is of order $M_{\rm p}R_{\rm p}v_{\rm esc}=(2GM_{\rm p}^3R_{\rm p})^{1/2}$, which is much larger than $S_{\rm init}$. Thus, we will assume the initial spin angular momentum of each planet is negligible in our analysis below.

\begin{figure}
    \begin{center}
	\includegraphics[width=0.6\columnwidth]{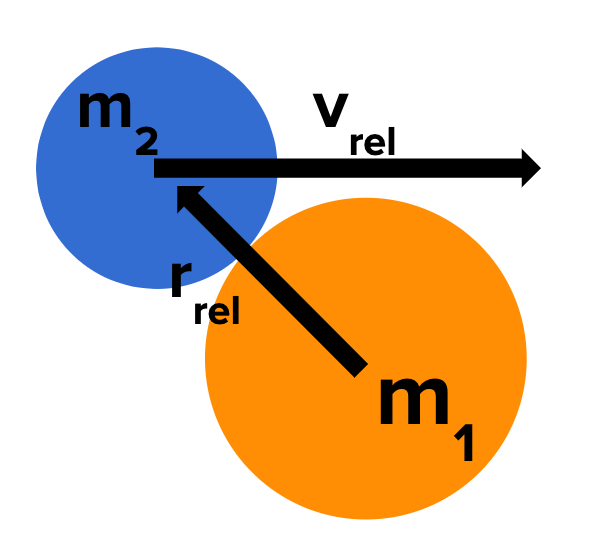}        
    \end{center}
    \caption{Geometry of collision between two planets. $\rrel$ and $\vrel$ are the relative position and velocity of between the two planets, taking $m_1$ as the reference.}
    \label{fig:setup-geo}
\end{figure}

With these assumptions, we can calculate the spin of the merger product as
\begin{equation}
    \Sm = \mu \rrel \times \vrel,
    \label{eq:Sm}
\end{equation}
where $\mu$ is the reduced mass of the two planets, $\rrel$ and $\vrel$ are the relative position and velocity between the planets at the moment of collision (see Fig.~\ref{fig:setup-geo}). The maximum value of spin is reached when $\rrel$ and $\vrel$ are perpendicular to each other. Taking $|\rrel|=R_1+R_2$ and $|\vrel|$ as the escape speed from $\rrel$, we expect that
\begin{equation}
    S_{\rm max} = \mu \sqrt{2G(m_1+m_2)(R_1+R_2)}
    \label{eq:Smax}
\end{equation}
is the maximum value of the spin angular momentum generated by collisions.

Since the mutual inclination between the initial planetary orbits is small, the merged object has an orbital angular momentum closely aligned with the normal unit vector $\nhat$ of the initial zero-inclination plane. The planetary obliquity, $\obl$, is then given by
\begin{equation}
    \cos\obl = \frac{\nhat \cdot \Sm}{|\Sm|}.
\end{equation}

\subsection{Fiducial results}
\label{sec:fid}

\begin{figure}
    \includegraphics[width=0.9\columnwidth]{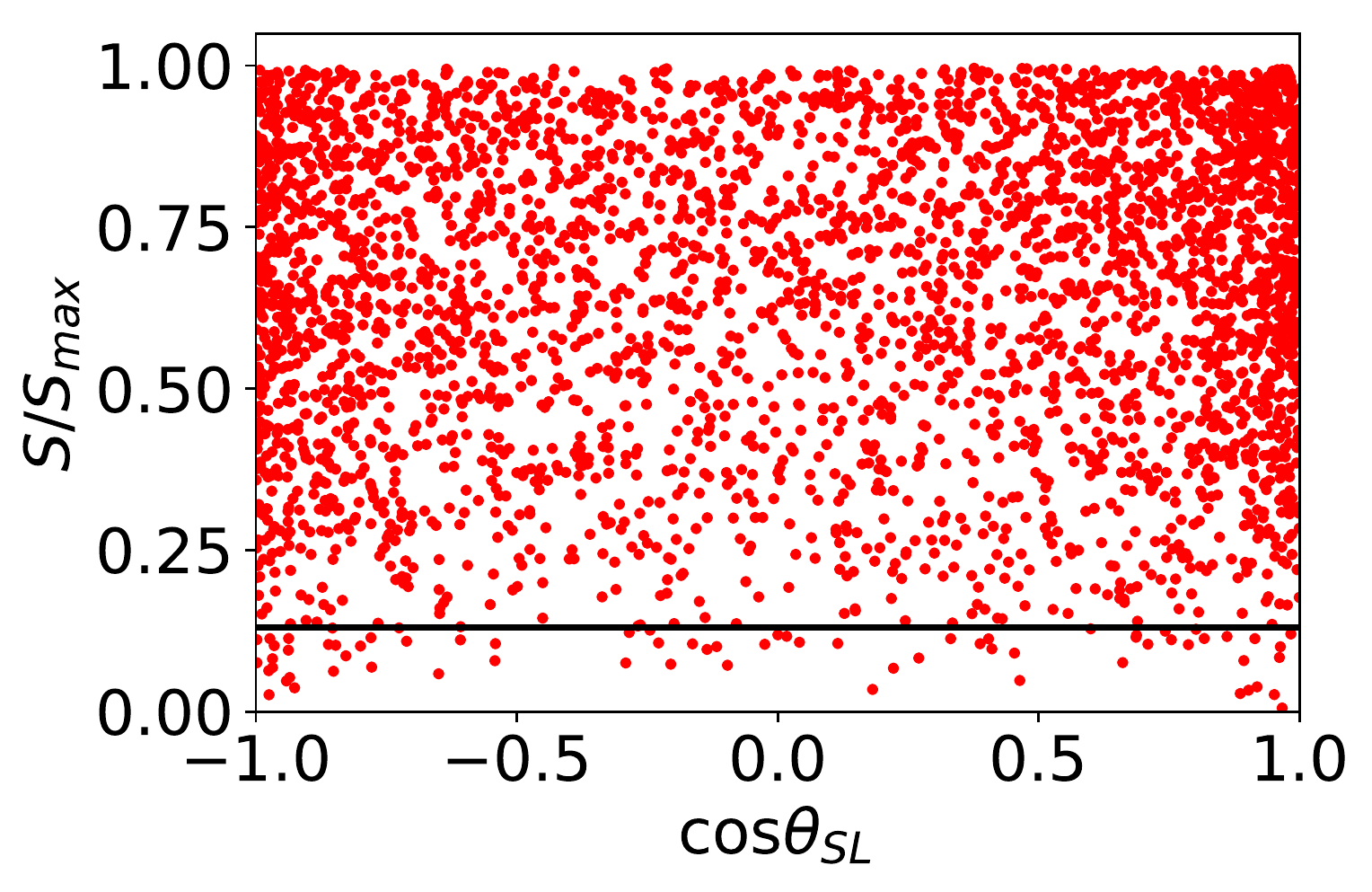}
    \caption{The spin and obliquity of the merger products found in our fiducial runs. The obliquity is displayed as $\cos\obl$ on the horizontal axis, and the spin is given in the unit of the maximum spin $S_{\rm max}$ (Eq.~\ref{eq:Smax}) on the vertical axis. The black line indicates the sum of the initial spins of the two planets (assuming each has $S_i=0.3I\Omega_{\rm break}$).}
\label{fig:res_fid_scatter}
\end{figure}

Fig.~\ref{fig:res_fid_scatter} shows the spin and obliquity of the merger products in our simulations. The values of spins are tightly bounded by the maximum given in Eq.~(\ref{eq:Smax}). Assuming both planets have an initial spin $0.3I\Omega_{\rm break}$ (see Section~\ref{sec:set-up}), the total initial spin is less than $0.13S_{\rm max}$. This means, in most cases, the relative orbital angular momentum at collision completely determines the final spin. Many merged objects have spins close to the maximum value, and they are strongly supported by rotation. Such object may lose a significant amount of angular momentum through deccretion and other processes, but we expect no change of its obliquity in the absence of further strong interactions with other planets.

\begin{figure}
	\includegraphics[width=0.9\columnwidth]{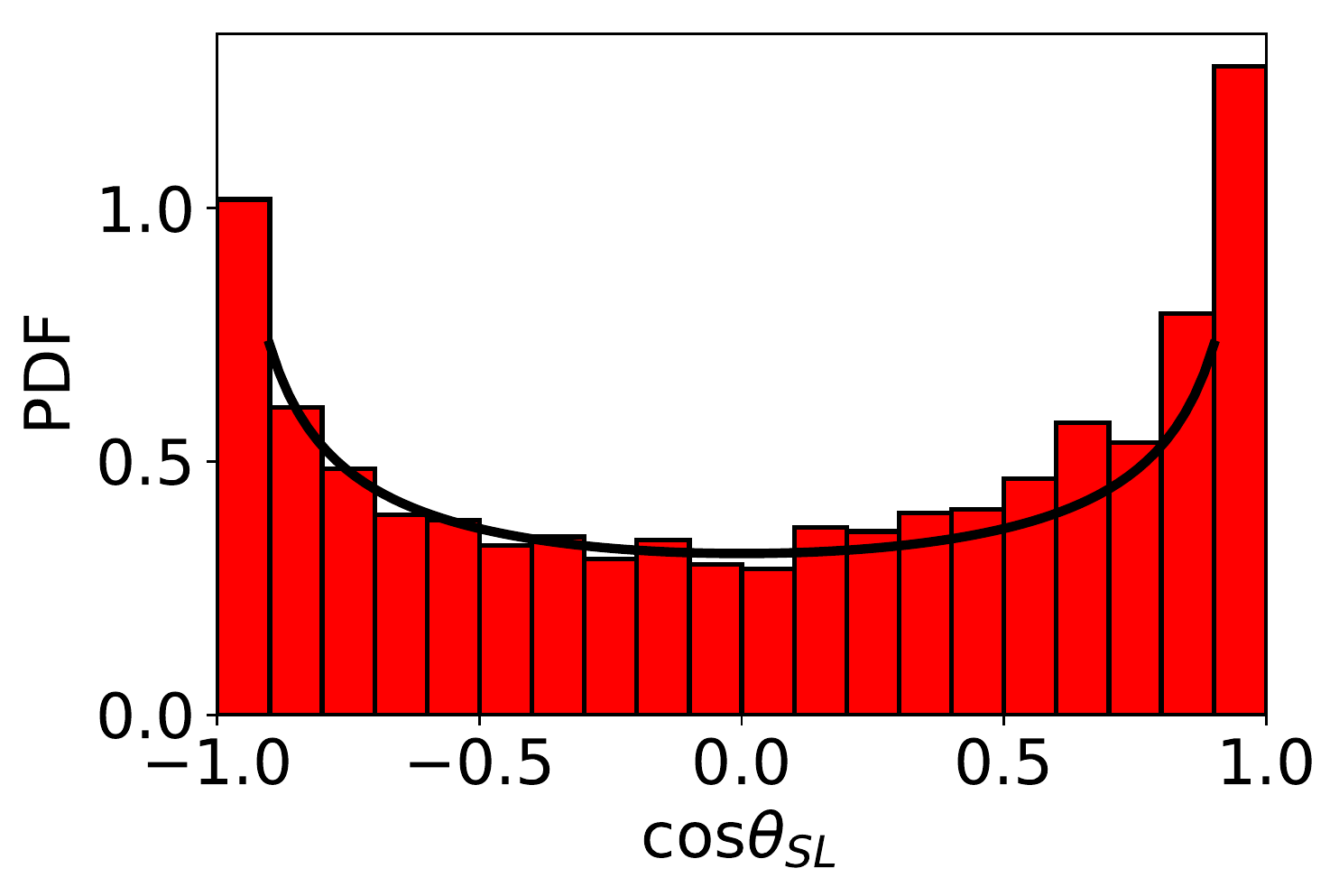}
	\includegraphics[width=0.9\columnwidth]{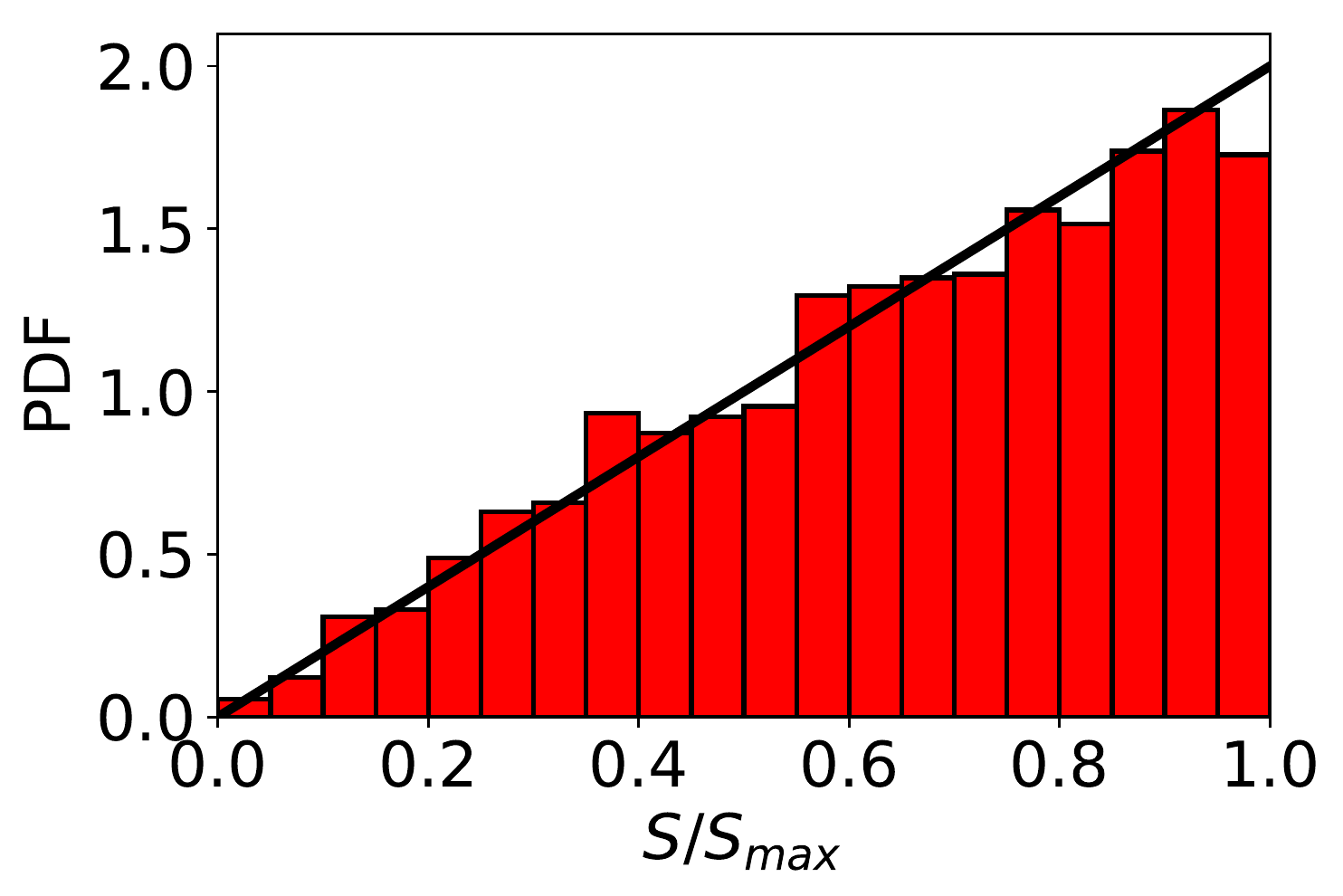}
    \caption{The obliquity (top) and spin (bottom) distribution of the merger products.The histograms are the numerical results found in our fiducial runs (with $a_1=1$~AU, $m_1 =2\MJ$, $m_2 = 1\MJ$, $R_1 = R_2 = \RJ$, ininitial inclination $i$ range $[0^{\circ},2^{\circ}]$). The black lines are the analytical distributions (Eqs.~\ref{eq:pdf-S} and~\ref{eq:pdf-obl}) derived in Section~\ref{sec:analytical}.}
    \label{fig:pdfs}
\end{figure}

Fig.~\ref{fig:pdfs} shows the marginalized distributions of obliquity and spin of the merged objects in our simulations. We also plot the analytical distributions derived in Section~\ref{sec:analytical}.

\begin{figure}
    \includegraphics[width=0.9\columnwidth]{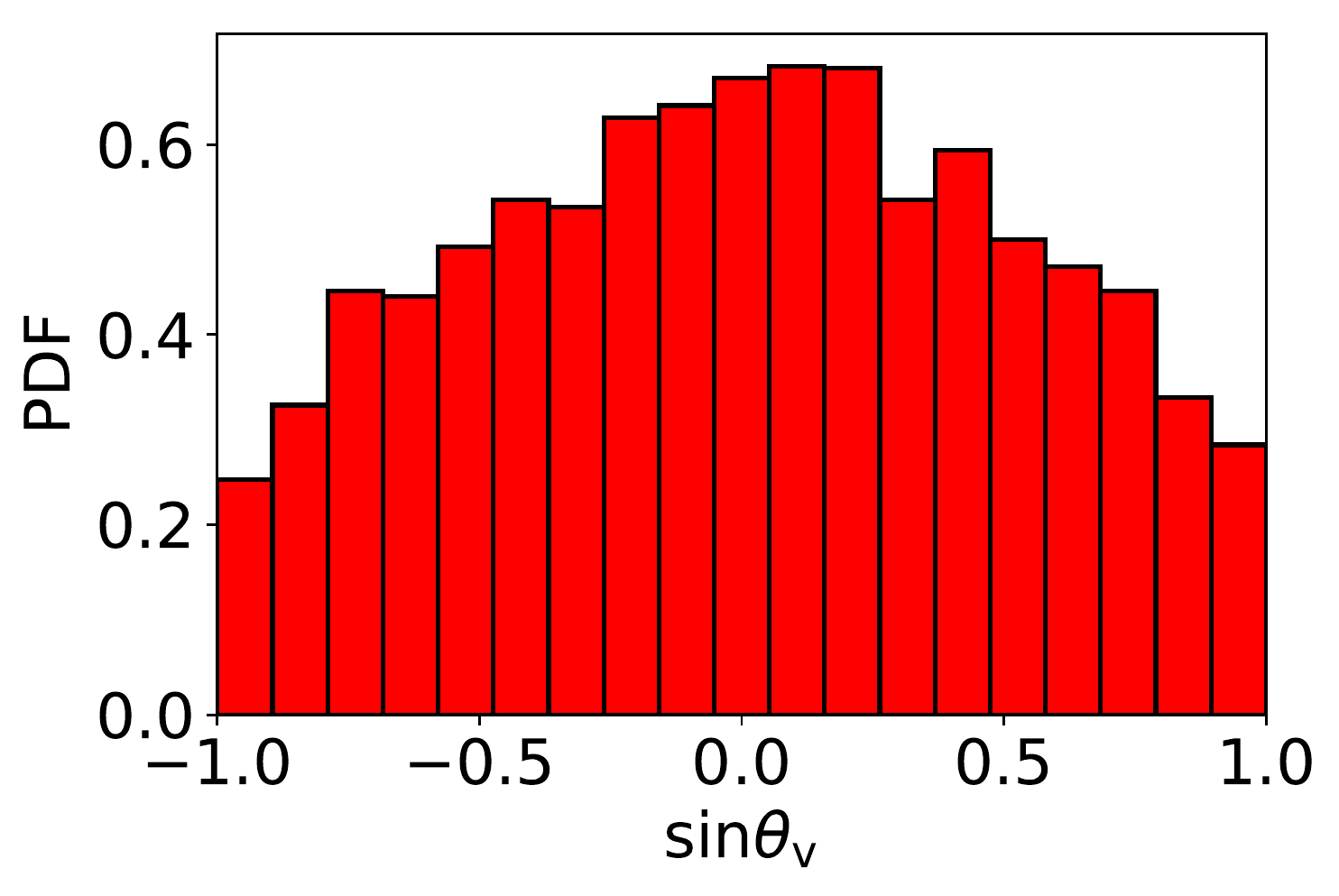}
    \includegraphics[width=0.9\columnwidth]{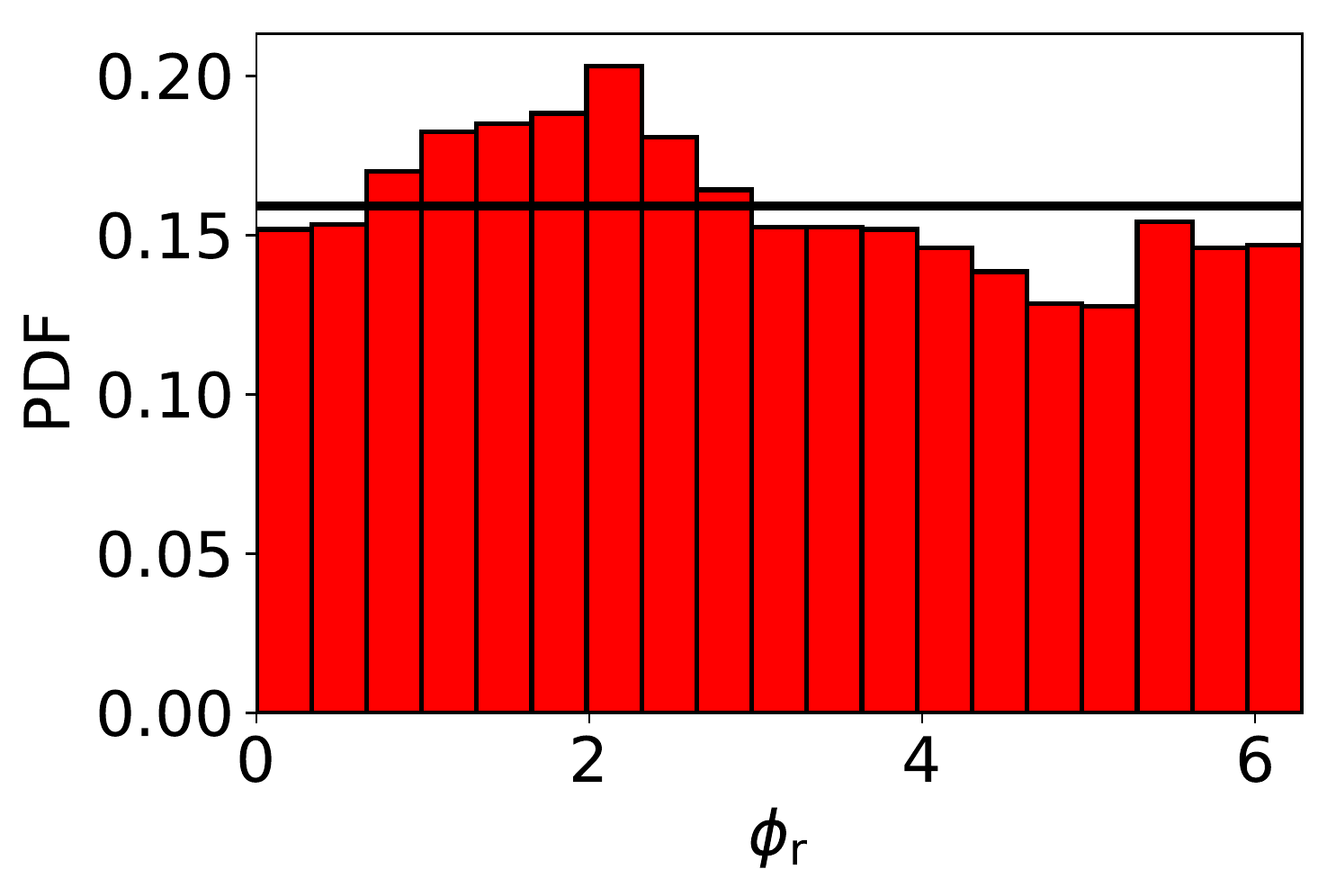}
    \includegraphics[width=0.9\columnwidth]{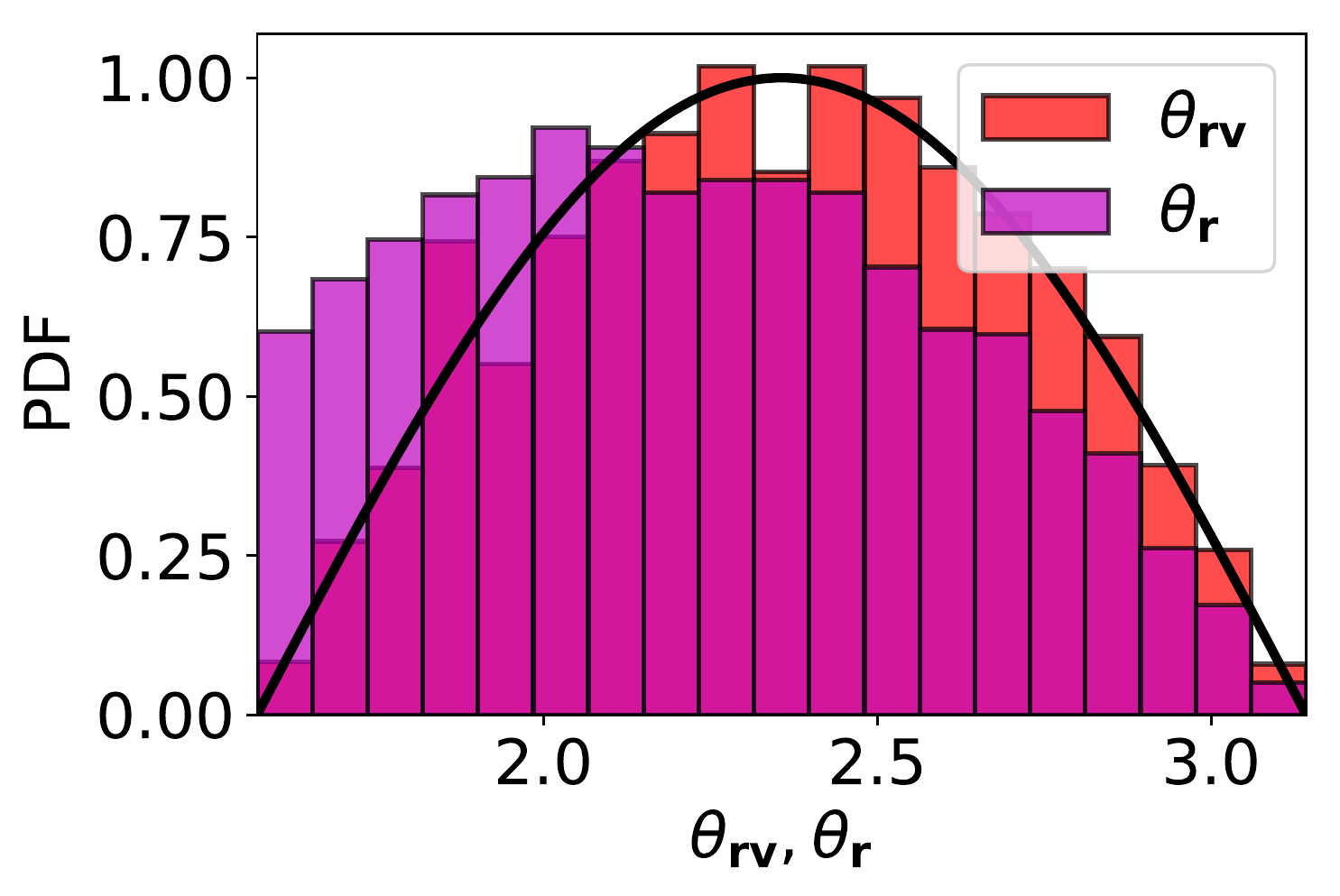}
    \caption{The distributions of $\theta_{\bf v}$ (defined in Eq.~\ref{eq:vcoor}),
    $\phir$ and $\thetar$ (defined in Eq.~\ref{eq:rcoor}). The distribution of $\theta_{\bf rv}$ (the angle between $\vrel$ and $\rrel$) is also shown for comparison. The histograms are the numerical results found in our fiducial runs, and the black line are the analytical distributions used in Section~\ref{sec:analytical}.}
    \label{fig:fid-geo}
\end{figure}

To gain some insight to the results, we investigate the geometry of collisions using Fig.~\ref{fig:fid-geo}. The normal of the orbital plane is denoted by $\nhat$. For each collision event, we decompose $\vrel$ into the vertical and ``in-plane'' components:
\begin{equation}
    \label{eq:vcoor}
    \vrel={\bf v}_{\rm proj} + (\vrel \cdot \nhat)\nhat =|\vrel|(\cos\theta_{\bf v} \vhat +\sin\theta_{\bf v} \nhat),
\end{equation}
where $\vhat = {\bf v}_{\rm proj} / |{\bf v}_{\rm proj}|$. The top panel of Fig.~\ref{fig:fid-geo} indicates that, at the moment of collision, $\vrel$ lies predominantly in the orbital plane. We define $\rhat = \vhat \times \nhat$ for each collision, and express $\rrel$ as
\begin{equation}
    \rrel = |\rrel|(\nhat \sin\thetar\cos\phir + \rhat \sin\thetar\sin\phir + \vhat \cos\thetar),
    \label{eq:rcoor}
\end{equation}
where $\thetar$ is the polar angle (with $\vhat$ pointing at the north pole) and $\phir$ is the azimuthal angle measured from $\nhat$. The middle panel of Fig.~\ref{fig:fid-geo} shows the distribution of $\phir$. The bottom panel shows the distribution of $\theta_{\bf rv}$ (the angle between $\vrel$ and $\rrel$) and $\thetar$ (the angle between ${\bf v}_{\rm proj}$ and $\rrel$). 

\section{Analytic Model for the Spin and Obliquity Distributions}
\label{sec:analytical}

\subsection{Analytical distributions}
\label{sec:geo}

Consider the moment when two planets collide (see Fig.~\ref{fig:setup-geo}).  We express $\rrel$ as in Eq.~(\ref{eq:rcoor}) and assume that the distribution of $\rphat = \rrel/|\rrel|$ is uniform inside an unit circle after being projected in the $\nhat-\rhat$ plane. Since the projection of $[\thetar,\thetar+d\thetar]\times[\phir,\phir+d\phir]$ in the $\nhat-\rhat$ plane has an area of $\sin\thetar d\sin\thetar d\phir$, the distributions for $\thetar$ and $\phir$ are
\begin{eqnarray}
    f_{\thetar} & = & -2\sin\thetar\cos\thetar, \label{eq:thetaphi-1}\\ 
    f_{\phir} & = & \frac{1}{2\pi}, \label{eq:thetaphi-2}
\end{eqnarray}
for $\thetar$ from $\pi/2$ to $\pi$ and $\phir$ from $0$ to $2\pi$. The two distributions are plotted as black lines in Fig.~\ref{fig:fid-geo}. We find an excellent matching between the analytical curves and the numerical results, except for a small asymmetry in the numerical $\phir$ distribution and a shift in the $\thetar$ distribution. This asymmetry implies a preferential alignment between the spin and the orbital angular momentum of the merger product over anti-alignment. The shift is due to the non-zero values of $\theta_{\bf v}$.

We further assume that the planets have a sufficiently low mutual inclination so that $\vrel \simeq {\bf v}_{\rm proj}$ (see the top panel of Fig.~\ref{fig:fid-geo}). From Eqs.~(\ref{eq:Sm})-(\ref{eq:Smax}), the spin of the merger product can be written as
\begin{equation}
    {\bf S} = S_{\rm max}(\nhat \sin\thetar\sin\phir - \rhat \sin\thetar\cos\phir).
\end{equation}
Thus, the spin magnitude and the obliquity are
\begin{eqnarray}
    & S/S_{\rm max} = \sin\thetar, \\
    & \cos\obl = \sin\phir . 
\end{eqnarray}
The distribution of $S/S_{\rm max}$ is then given by 
\begin{equation}
    f_{S/S_{\rm max}} = f_{\thetar} \left |\frac{d(S/S_{\rm max})}{d\thetar}\right |^{-1} = 2\sin\thetar = 2\frac{S}{S_{\rm max}}.
    \label{eq:pdf-S}
\end{equation}
The distribution of $\cos\obl$ is
\begin{equation}
    f_{\cos\obl} = 2f_{\phir} \left|\frac{d\cos\obl}{d\phir}\right|^{-1}= \frac{1}{\pi |\cos\phir|}
    = \frac{1}{\pi}\frac{1}{\sqrt{1-\cos^2\obl}},
    \label{eq:pdf-obl}
\end{equation}
where the factor of 2 comes from the fact that the inverse function of $\sin\phir$ is double-valued for $\phir$ from $0$ to $2\pi$. The two analytical distributions are plotted as black lines in Fig.~\ref{fig:pdfs}, showing excellent agreement to the numerical results.

\subsection{Validity and Limitation}
\label{sec:limitation}

There are mainly two limitations to our analytical distributions. The first occurs when the initial mutual inclination between the
planetary orbits is too small. Our assumption of the $(\thetar,\phir)$
distribution (Eqs.~\ref{eq:thetaphi-1} and~\ref{eq:thetaphi-2}) requires equal accessibility to
any points in the area perpendicular to $-\vhat$. This is possible only when
the initial mutual inclination between the two planetary orbits is at least a few times
larger than $R_{\rm p}/a_1$.

The second limitation occurs when the initial mutual inclination is too large. In Section~\ref{sec:geo} we have assumed $|\vrel\cdot\nhat|\ll |{\bf v}_{\rm proj}|$, or $\theta_v\simeq 0$. Suppose the outer planet (initially at semimajor axis $a_2$) moves to the inner planet (at $a_1$) and enters the mutual Hill sphere.  At this point, their relative velocity in orbital plane can be estimated as $v_\parallel \sim \sqrt{GM_*a_2}/a_1-\sqrt{GM_*/a_1}\simeq (\Delta a/2a_1)\sqrt{GM_*/a_1}$, where $\Delta a=a_2-a_1$. On the other hand, the vertical velocity difference almost solely comes from the mutual inclination, $v_{\perp} \sim \sqrt{GM_*/a_1} \sin i$. After entering the mutual Hill sphere, the relative motion between the two planets is governed by their mutual gravitational attraction, and the orientation of the ``binary'' relative to the original orbital plane remains approximately constant. Thus, at collision, the inclination angle ($\theta_{\bf v}$) of $\vrel$ relative to the original orbital plane is given by $\tan\theta_{\bf v} \sim (2a_1/\Delta a)\sin i$. The condition $\theta_{\bf v}\ll 1$ is equivalent to $\sin i \ll \Delta a/(2a_1)$. 

When $\theta_{\bf v}$ is non-negligible, we expect that, instead of the $\nhat$-$\rhat$-plane, $\rphat$ is uniform in the plane normal to the actual relative velocity $\vrel$. The distribution of $|\Sm|$ would be similar to that derived in Section~\ref{sec:geo} (since the change from $\vhat$ to ${\hat{\bf v}}_{\rm rel}$ amounts to a simple rotation of the coordinate system). However, the obliquity becomes
\begin{equation}
    \cos\obl = \cos\theta_v\sin\thetar\sin\phir/|\Sm|.
\end{equation}
A finite $\theta_v$ tends to reduce $|\cos\obl |$, with the corresponding change in the distribution of
$\cos\obl$.

In summary, we expect that the analytical distribution of $\obl$ (Eq.~\ref{eq:pdf-obl}) to be valid when
\begin{equation}
    \frac{R_{\rm p}}{a_1} \ll \sin i \ll \frac{\Delta a}{2a_1} \simeq \frac{K}{2} \left( \frac{m_1+m_2}{3M_*} \right)^{1/3},
    \label{eq:limit-obl}
\end{equation}
where we have used $\Delta a = KR_{\rm H,mut}$ (see Eq.~\ref{eq:RH}). On the other hand, the analytical distribution of $S$ (Eq.~\ref{eq:pdf-S}) is valid when \begin{equation}
    \frac{R_{\rm p}}{a_1} \ll \sin i.
    \label{eq:limit-S}
\end{equation}
For the fiducial numerical simulations presented in Section~\ref{sec:simulation}, these conditions are well satisfied: an initial inclination of $2^{\circ}$ corresponds to $73 \RJ/(1~{\rm AU}) =0.035$, and is much less than $\Delta a/(2a_1)=0.14$. So it is not surprising that $S/S_{\rm max}$ and $\obl$ follow the analytical distributions very well.


\section{Parameter Studies}
\label{sec:parameter}

We perform parameter studies by carrying out simulations with different initial semi-major axis, mutual inclination, and planet radius to test the validity and limitations of our fiducial results (Section~\ref{sec:simulation}) and analytical formulae (Section~\ref{sec:analytical}).

\subsection{Initial semi-major axis}
\begin{figure*}
    \begin{center}
    \includegraphics[width=1.8\columnwidth]{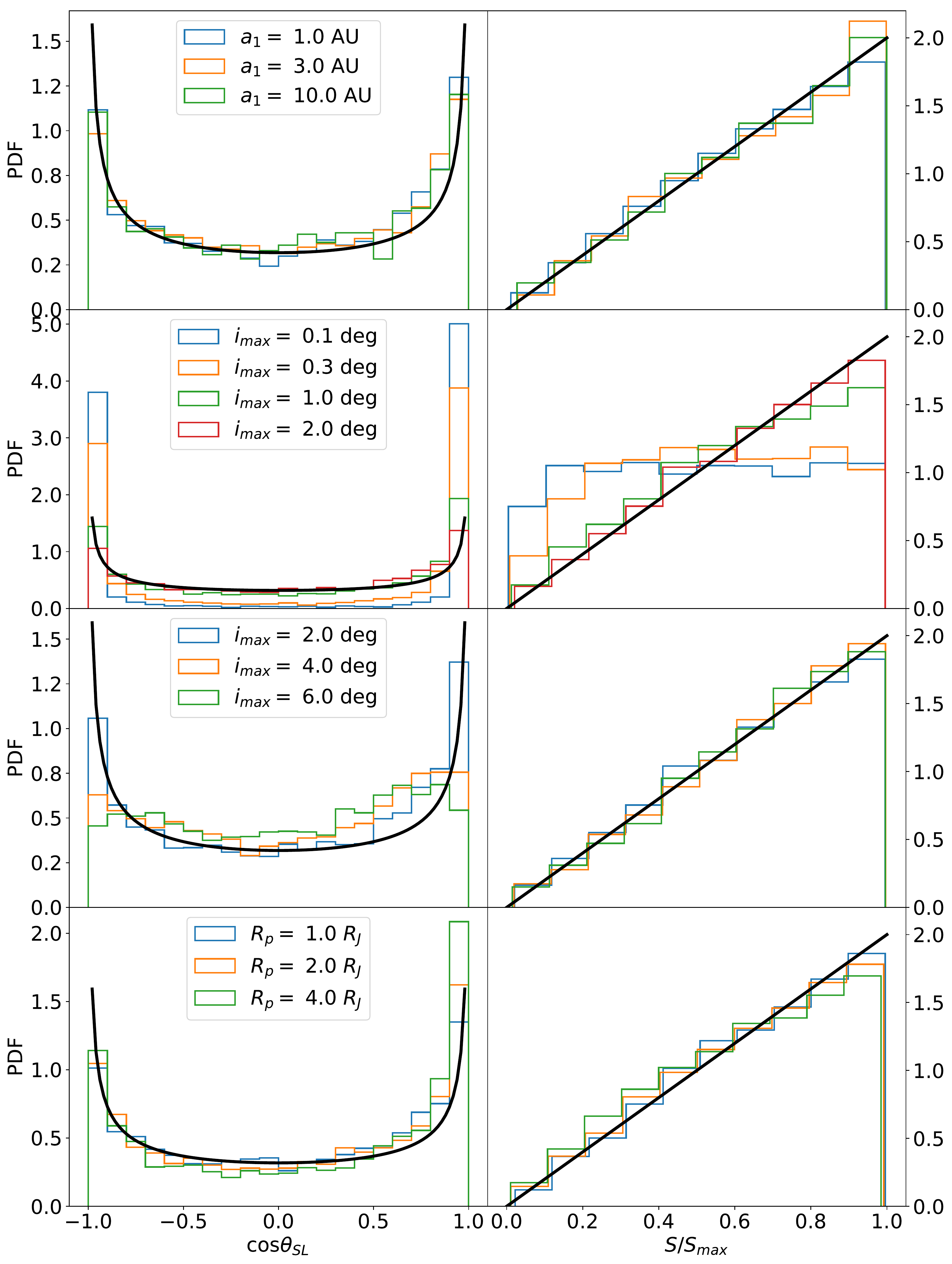}	        
    \end{center}
    \caption{Same as Fig.~\ref{fig:pdfs}, except for different values of the parameter used in the simulations. Top row: initial $a_1$; middle two rows: $i_{\rm max}$ (the initial inclinations of both planets are sampled from $[0,i_{\rm max}]$); bottom row: and the planet radius $R_{\rm p}$ as labeled.}
    \label{fig:pdf_par}
\end{figure*}

Note that varying $a_1$ also changes $a_2$ according to Eq.~(\ref{eq:a1a2}). Increasing $a_1$ makes ejections more likely than collisions as the outcomes of planetary scatterings see \citep[see][]{Li2020}. It also makes the systems safer from the lower limit in Eqs.~(\ref{eq:limit-obl})-(\ref{eq:limit-S}). As expected, the top row of Fig.~\ref{fig:pdf_par} shows that the distributions of the spin and obliquity for different $a_1$ values are the same as the fiducial results.

\subsection{Initial inclination}

We investigate the effect of the initial inclination by changing the upper limit of the initial inclination, $2\dego$, to different values of $i_{\rm max}$.

The second row of Fig.~\ref{fig:pdf_par} shows the results for small $i_{\rm max}$'s. Note that $i_{\rm max}=0.1\dego$ and $2.0\dego$ correspond to $3.7$ and $73 (\RJ/a_1)$, respectively. For small $i_{\rm max}$, we expect the analytical distribution
$f_{\thetar}$ and $f_{\phir}$ (Eqs.~\ref{eq:thetaphi-1} and~\ref{eq:thetaphi-2}) to fail (see Eqs.~\ref{eq:limit-obl} and~\ref{eq:limit-S}). The numerical result shows that obliquities are more concentrated around $0\dego$ or $180\dego$ for $i_{\rm max}=0.1\dego$ and $0.3\dego$, and the distribution of the spin magnitude tends to be more uniform.

The third row of Fig.~\ref{fig:pdf_par} shows the results for $i_{\rm max}$ equal to a few degrees. In this range, the simulated systems are safe from the lower inclination limit of Eqs.~(\ref{eq:limit-obl})-(\ref{eq:limit-S}). As expected (Section~\ref{sec:limitation}), the analytical distribution for $S/S_{\rm max}$ matches the numerical results well. However, as $i_{\rm max}$ increases to more than $4\dego$, the numerical obliquity distribution starts to deviate from the analytical expression. Using $\tan\theta_{\bf v} \sim (2a_1/\Delta a)\sin i$, we find that $\tan \theta_v \sim 0.24$, $0.49$, and $0.73$ for $i=2\dego$, $4\dego$, and $6\dego$, respectively. Hence, for $i_{\rm max}=4\dego$ and $6\dego$, our analytic $\cos\obl$ distribution becomes inaccurate.

\subsection{Size of planet}

Varying the size of the planets can change the branching ratio of mergers vs ejections (see Li et al.~2020), but does not cause any change to our results concerning the spin and obliquity distributions, as long as the condition $R_{\rm p}/a_1\ll i_{\rm max}$ is satisfied. Note that the value of $R_{\rm p}$ is irrelevant to our assumption of $\theta_v\simeq 0$ (i.e. ${\bf v}_{\rm rel}$ is in the orbital plane). The bottom row of Fig.~\ref{fig:pdf_par} shows the results. The magnitude of the spin is normalized by different $S_{\rm max}$ according to the planet's radius. As expected, the plots are similar to the fiducial results (Fig.~\ref{fig:pdfs}).


\section{Summary and Discussion}
\label{sec:summary}
We have carried out a suite of numerical simulations of the dynamical evolution of two giant planets, initially in quasi-circular unstable orbits, to determine the distributions of spin and obliquity of the planet merger products. While many previous works have studied giant planet scatterings, our work is the first (as far was we know) to systematically determine the spin property of the planet mergers. Based on our recent work on the hydrodynamics of giant planet collisions \citep{Li2020}, we assume that two colliding planets always merge into a bigger one with no mass and angular momentum losses. For reasonable initial (pre-merger) rotations of the planets, the spin angular momentum of the merger product is dominated by the relative orbital angular momentum of the colliding planets at contact.

Our most important finding is displayed in Fig.~\ref{fig:pdfs}, showing the distributions of $\cos\obl$ (where $\obl$ is the obliquity or spin-orbit misalignment angle) and spin $S$ (in units of the maximum possible value) of the planet merger products in our fiducial runs. We develop a simple model and show that these distributions are well described by the analytical expressions (Eqs.~\ref{eq:pdf-S} and \ref{eq:pdf-obl})
\begin{equation}
  f_{S/S_{\rm max}}={2S/S_{\rm max}},\quad
  f_{\cos\theta_{\rm SL}}=\frac{1}{\pi\sqrt{1-\cos^2\theta_{\rm SL}}}.
\end{equation}
In addition, we carry out parameter studies to explore the validity of these distributions under various conditions (Section 4). The key is the initial mutual inclination $\Delta i$ of the planetary orbits, which is limited by $i_{\rm max}$ in our parameter studies, in relation to the size $R_{\rm p}$ and initial spacing $\Delta a$ of the planets (see Eqs.~\ref{eq:limit-obl} and~\ref{eq:limit-S}). We find that the analytic distribution of $S$ works well as long as $\Delta i$ is much greater than $R_{\rm p}/a$ (Eq.~\ref{eq:limit-S}), while the analytic distribution of $\cos\obl$ further requires that $\Delta i$ be much less than $\Delta a/(2a)$ (Eq.~\ref{eq:limit-obl}) -- when this condition is not satisfied, a more uniform distribution of $\cos\obl$ is obtained (see the third row of Fig.~\ref{fig:pdf_par}).

A possible caveat of this study is that we have neglected tidal effects in close planet-planet encounters. Numerical simulations of planet  scatterings including hydrodynamical effects show that only the collision vs ejection branching ratios are affected, while the spin and obliquity distributions of the merger products are mostly unchanged by the tidal effects \citep{Li2020}.

While in this paper we have focused on mergers of giant planets, we expect that similar results may hold for mergers of smaller planets such as super-Earths and mini-Neptunes. We simply need to compare $\Delta i$, $R_{\rm p}/a$ and $\Delta a/(2a)$ to determine the regimes of validity of our analytical and numerical results.

Overall, our study shows that planetary mergers predominantly produce rapid rotating objects. These objects are rotationally supported, and are obviously quite different from the ``usual'' planets. Their spins may undergo further evolution, so the present-day distribution of $S$ could well be different from what is predicted in this paper. However, we expect the obliquity and its distribution to be more ``permenant''.  Observational search for the merger signatures in the form of spin and obliquity, for various types of planets, will be valuable in constraining the dynamical history of planetary systems.

\acknowledgments

DL thanks the Dept. of Astronomy and the Miller Institute for Basic Science at UC Berkeley for hospitality while part of this work was carried out. This work has been supported in part by the NSF grant AST-17152 and NASA grant 80NSSC19K0444. \\



\software{Matplotlib \citep{Hunter2007}, NumPy \citep{Walt2011}, REBOUND \citep{Rein2012}
          }



\bibliographystyle{aasjournal}

\begin{thebibliography}{}
\expandafter\ifx\csname natexlab\endcsname\relax\def\natexlab#1{#1}\fi
\providecommand{\url}[1]{\href{#1}{#1}}
\providecommand{\dodoi}[1]{doi:~\href{http://doi.org/#1}{\nolinkurl{#1}}}
\providecommand{\doeprint}[1]{\href{http://ascl.net/#1}{\nolinkurl{http://ascl.net/#1}}}
\providecommand{\doarXiv}[1]{\href{https://arxiv.org/abs/#1}{\nolinkurl{https://arxiv.org/abs/#1}}}

\bibitem[{Adams \& Laughlin(2003)}]{Adams2003}
Adams, F.~C., \& Laughlin, G. 2003, Icarus, 163, 290,
  \dodoi{10.1016/s0019-1035(03)00081-2}

\bibitem[{Agnor \& Asphaug(2004)}]{Agnor2004}
Agnor, C., \& Asphaug, E. 2004, The Astrophysical Journal, 613, L157,
  \dodoi{10.1086/425158}

\bibitem[{Anderson {et~al.}(2020)Anderson, Lai, \& Pu}]{Anderson2020}
Anderson, K.~R., Lai, D., \& Pu, B. 2020, Monthly Notices of the Royal
  Astronomical Society, 491, 1369, \dodoi{10.1093/mnras/stz3119}

\bibitem[{Asphaug {et~al.}(2006)Asphaug, Agnor, \& Williams}]{Asphaug2006}
Asphaug, E., Agnor, C.~B., \& Williams, Q. 2006, Nature, 439, 155,
  \dodoi{10.1038/nature04311}

\bibitem[{Barnes \& Fortney(2003)}]{Barnes2003}
Barnes, J.~W., \& Fortney, J.~J. 2003, The Astrophysical Journal, 588, 545,
  \dodoi{10.1086/373893}

\bibitem[{Batygin(2018)}]{Batygin2018}
Batygin, K. 2018, The Astronomical Journal, 155, 178,
  \dodoi{10.3847/1538-3881/aab54e}

\bibitem[{{Benz} {et~al.}(1989){Benz}, {Slattery}, \& {Cameron}}]{Benz1989}
{Benz}, W., {Slattery}, W.~L., \& {Cameron}, A.~G.~W. 1989, Meteoritics, 24,
  251

\bibitem[{Bryan {et~al.}(2017)Bryan, Benneke, Knutson, Batygin, \&
  Bowler}]{Bryan2017}
Bryan, M.~L., Benneke, B., Knutson, H.~A., Batygin, K., \& Bowler, B.~P. 2017,
  Nature Astronomy, 2, 138, \dodoi{10.1038/s41550-017-0325-8}

\bibitem[{{Bryan} {et~al.}(2020){Bryan}, {Chiang}, {Bowler}, {Morley},
  {Millholland}, {Blunt}, {Ashok}, {Nielsen}, {Ngo}, {Mawet}, \&
  {Knutson}}]{Bryan2020}
{Bryan}, M.~L., {Chiang}, E., {Bowler}, B.~P., {et~al.} 2020, \aj, 159, 181,
  \dodoi{10.3847/1538-3881/ab76c6}

\bibitem[{Chambers {et~al.}(1996)Chambers, Wetherill, \& Boss}]{Chambers1996}
Chambers, J., Wetherill, G., \& Boss, A. 1996, Icarus, 119, 261,
  \dodoi{10.1006/icar.1996.0019}

\bibitem[{Chatterjee {et~al.}(2008)Chatterjee, Ford, Matsumura, \&
  Rasio}]{Chatterjee2008}
Chatterjee, S., Ford, E.~B., Matsumura, S., \& Rasio, F.~A. 2008, The
  Astrophysical Journal, 686, 580, \dodoi{10.1086/590227}

\bibitem[{Dones \& Tremaine(1993)}]{Dones1993}
Dones, L., \& Tremaine, S. 1993, Icarus, 103, 67,
  \dodoi{10.1006/icar.1993.1059}

\bibitem[{Ford {et~al.}(2001)Ford, Havlickova, \& Rasio}]{Ford2001}
Ford, E.~B., Havlickova, M., \& Rasio, F.~A. 2001, Icarus, 150, 303,
  \dodoi{10.1006/icar.2001.6588}

\bibitem[{Ford \& Rasio(2008)}]{Ford2008}
Ford, E.~B., \& Rasio, F.~A. 2008, The Astrophysical Journal, 686, 621,
  \dodoi{10.1086/590926}

\bibitem[{Ginzburg \& Chiang(2019)}]{Ginzburg2019}
Ginzburg, S., \& Chiang, E. 2019, Monthly Notices of the Royal Astronomical
  Society: Letters, 491, L34, \dodoi{10.1093/mnrasl/slz164}

\bibitem[{Gladman(1993)}]{Gladman1993}
Gladman, B. 1993, Icarus, 106, 247, \dodoi{10.1006/icar.1993.1169}

\bibitem[{Hamilton \& Ward(2004)}]{Hamilton2004}
Hamilton, D.~P., \& Ward, W.~R. 2004, The Astronomical Journal, 128, 2510,
  \dodoi{10.1086/424534}

\bibitem[{Hunter(2007)}]{Hunter2007}
Hunter, J.~D. 2007, Computing in Science {\&} Engineering, 9, 90,
  \dodoi{10.1109/mcse.2007.55}

\bibitem[{Juri{\'{c}} \& Tremaine(2008)}]{Juric2008}
Juri{\'{c}}, M., \& Tremaine, S. 2008, The Astrophysical Journal, 686, 603,
  \dodoi{10.1086/590047}

\bibitem[{Korycansky {et~al.}(1990)Korycansky, Bodenheimer, Cassen, \&
  Pollack}]{Korycansky1990}
Korycansky, D., Bodenheimer, P., Cassen, P., \& Pollack, J. 1990, Icarus, 84,
  528, \dodoi{10.1016/0019-1035(90)90051-a}

\bibitem[{Leinhardt \& Stewart(2011)}]{Leinhardt2011}
Leinhardt, Z.~M., \& Stewart, S.~T. 2011, The Astrophysical Journal, 745, 79,
  \dodoi{10.1088/0004-637x/745/1/79}

\bibitem[{Li {et~al.}(2020)Li, Lai, \& Anderson}]{Li2020}
Li, J., Lai, D., \& Anderson, K.~R. 2020, Monthly Notices of the Royal
  Astronomical Society, Submitted

\bibitem[{Lin \& Ida(1997)}]{Lin1997}
Lin, D. N.~C., \& Ida, S. 1997, The Astrophysical Journal, 477, 781,
  \dodoi{10.1086/303738}

\bibitem[{Lissauer {et~al.}(1997)Lissauer, Berman, Greenzweig, \&
  Kary}]{Lissauer1997}
Lissauer, J.~J., Berman, A.~F., Greenzweig, Y., \& Kary, D.~M. 1997, Icarus,
  127, 65, \dodoi{10.1006/icar.1997.5689}

\bibitem[{Millholland \& Batygin(2019)}]{Millholland2019}
Millholland, S., \& Batygin, K. 2019, The Astrophysical Journal, 876, 119,
  \dodoi{10.3847/1538-4357/ab19be}

\bibitem[{Morbidelli {et~al.}(2012)Morbidelli, Tsiganis, Batygin, Crida, \&
  Gomes}]{Morbidelli2012}
Morbidelli, A., Tsiganis, K., Batygin, K., Crida, A., \& Gomes, R. 2012,
  Icarus, 219, 737, \dodoi{10.1016/j.icarus.2012.03.025}

\bibitem[{Nagasawa \& Ida(2011)}]{Nagasawa2011}
Nagasawa, M., \& Ida, S. 2011, The Astrophysical Journal, 742, 72,
  \dodoi{10.1088/0004-637x/742/2/72}

\bibitem[{Petrovich {et~al.}(2014)Petrovich, Tremaine, \&
  Rafikov}]{Petrovich2014}
Petrovich, C., Tremaine, S., \& Rafikov, R. 2014, The Astrophysical Journal,
  786, 101, \dodoi{10.1088/0004-637x/786/2/101}

\bibitem[{Rasio \& Ford(1996)}]{Rasio1996}
Rasio, F.~A., \& Ford, E.~B. 1996, Science, 274, 954,
  \dodoi{10.1126/science.274.5289.954}

\bibitem[{Rein \& Liu(2012)}]{Rein2012}
Rein, H., \& Liu, S.-F. 2012, Astronomy {\&} Astrophysics, 537, A128,
  \dodoi{10.1051/0004-6361/201118085}

\bibitem[{Rein \& Spiegel(2014)}]{Rein2014}
Rein, H., \& Spiegel, D.~S. 2014, Monthly Notices of the Royal Astronomical
  Society, 446, 1424, \dodoi{10.1093/mnras/stu2164}

\bibitem[{Rogoszinski \& Hamilton(2020)}]{Rogoszinski2020}
Rogoszinski, Z., \& Hamilton, D.~P. 2020, The Astrophysical Journal, 888, 60,
  \dodoi{10.3847/1538-4357/ab5d35}

\bibitem[{Safronov \& Zvjagina(1969)}]{Safronov1969}
Safronov, V., \& Zvjagina, E. 1969, Icarus, 10, 109,
  \dodoi{10.1016/0019-1035(69)90013-x}

\bibitem[{Schwartz {et~al.}(2016)Schwartz, Sekowski, Haggard, Pall{\'{e}}, \&
  Cowan}]{Schwartz2016}
Schwartz, J.~C., Sekowski, C., Haggard, H.~M., Pall{\'{e}}, E., \& Cowan, N.~B.
  2016, Monthly Notices of the Royal Astronomical Society, 457, 926,
  \dodoi{10.1093/mnras/stw068}

\bibitem[{Seager \& Hui(2002)}]{Seager2002}
Seager, S., \& Hui, L. 2002, The Astrophysical Journal, 574, 1004,
  \dodoi{10.1086/340994}

\bibitem[{Snellen {et~al.}(2014)Snellen, Brandl, de~Kok, Brogi, Birkby, \&
  Schwarz}]{Snellen2014}
Snellen, I. A.~G., Brandl, B.~R., de~Kok, R.~J., {et~al.} 2014, Nature, 509,
  63, \dodoi{10.1038/nature13253}

\bibitem[{Stewart \& Leinhardt(2012)}]{Stewart2012}
Stewart, S.~T., \& Leinhardt, Z.~M. 2012, The Astrophysical Journal, 751, 32,
  \dodoi{10.1088/0004-637x/751/1/32}

\bibitem[{{Su} \& {Lai}(2020)}]{Su2020}
{Su}, Y., \& {Lai}, D. 2020, arXiv e-prints, arXiv:2004.14380.
\newblock \doarXiv{2004.14380}

\bibitem[{Takata \& Stevenson(1996)}]{Takata1996}
Takata, T., \& Stevenson, D.~J. 1996, Icarus, 123, 404,
  \dodoi{10.1006/icar.1996.0167}

\bibitem[{Tremaine(1991)}]{Tremaine1991}
Tremaine, S. 1991, Icarus, 89, 85, \dodoi{10.1016/0019-1035(91)90089-c}

\bibitem[{van~der Walt {et~al.}(2011)van~der Walt, Colbert, \&
  Varoquaux}]{Walt2011}
van~der Walt, S., Colbert, S.~C., \& Varoquaux, G. 2011, Computing in Science
  {\&} Engineering, 13, 22, \dodoi{10.1109/mcse.2011.37}

\bibitem[{Vokrouhlick{\'{y}} \& Nesvorn{\'{y}}(2015)}]{Vokrouhlicky2015}
Vokrouhlick{\'{y}}, D., \& Nesvorn{\'{y}}, D. 2015, The Astrophysical Journal,
  806, 143, \dodoi{10.1088/0004-637x/806/1/143}

\bibitem[{Ward \& Hamilton(2004)}]{Ward2004}
Ward, W.~R., \& Hamilton, D.~P. 2004, The Astronomical Journal, 128, 2501,
  \dodoi{10.1086/424533}

\end{thebibliography}



\end{document}